\documentclass[universe,conferencereport,accept,moreauthors,pdftex,10pt,a4paper]{mdpi} 

\usepackage{microtype}
\usepackage{soul}
\firstpage{1} 
\articlenumber{44}
\doinum{10.3390/universe3020044}
\pubvolume{2}
\pubyear{2017}
\copyrightyear{2017}
\externaleditor{Academic Editors: Mariusz P. D{\c a}browski, Manuel Kr{\"a}mer and Vincenzo Salzano}
\history{Received: 9 January 2017; Accepted: 3 May 2017; Published: 9 May 2017}

\newcommand{\Hi}{H\,{\sc i}}

\newcommand{\Di}{D\,{\sc i}}
\newcommand{\DtH}{\Di/\Hi{}}

\newcommand{\Oi}{O\,{\sc i}}

\newcommand{\kms}{\, \mathrm{km\cdot s}^{-1} }
 					
\newcommand{\cm}{\, \mathrm{cm} }

\newcommand{\Lya}{Lyman\,$\alpha$}

\newcommand{\vpfit}{\texttt{VPFIT}}

\newcommand{\Planck}{the {\sl Planck Surveyor}} 

\newcommand\tabref[1]{%
Table~\ref{tab:#1}}

\newcommand\figref[1]{%
Figure~\ref{fig:#1}}

\newcommand\secref[1]{%
Section~\ref{sec:#1}}

\usepackage[dvipsnames]{xcolor}

\Title{Nucleosynthesis Predictions and High-Precision Deuterium Measurements}

\Author{Signe Riemer-S\o{}rensen * and Espen Sem Jenssen}
\AuthorNames{Signe Riemer-S\o{}rensen and Espen Sem Jenssen}

\address[1]{%
Institute of Theoretical Astrophysics, The University of Oslo, Boks 1072 Blindern, NO-0316 Oslo, Norway;   EspenJenssen@hotmail.com\\
}

\corres{\hspace{-1em}Correspondence: signe.riemer-sorensen@astro.uio.no}

\abstract{Two new high-precision measurements of the deuterium abundance from absorbers along the line of sight to the quasar PKS1937--1009 were presented. The absorbers have lower neutral hydrogen column densities (N(HI) $\approx$ 18\,cm$^{-2}$) than for previous high-precision measurements, boding well for further extensions of the sample due to the plenitude of low column density absorbers. The total high-precision sample now consists of 12 measurements with a weighted average deuterium abundance of D/H = $2.55\pm0.02\times10^{-5}$. The sample does not favour a dipole similar to the one detected for the fine structure constant. The increased precision also calls for improved nucleosynthesis predictions. For that purpose we have updated the public AlterBBN code including new reactions, updated nuclear reaction rates, and the possibility of adding new physics such as dark matter. The standard Big Bang Nucleosynthesis prediction of D/H = $2.456\pm0.057\times10^{-5}$ is consistent with the observed value within 1.7 standard deviations.}


\begin{document}


\section{Introduction}
Consisting only of a proton and a neutron, deuterium is the simplest element, and consequently it was the first one to form in the early universe. The nucleosynthesis of the light elements such as deuterium, helium and lithium took place during the first few minutes after Big Bang when the universe was still opaque, and today we can only probe the conditions indirectly through the resulting abundances. This requires high precision measurements combined with reliable predictions for the abundances. In \secref{background} we provide a short overview of the nucleosynthesis, before we present new high precision measurements of the deuterium abundance in \secref{deuteriumobs} and a new code for performing predictions in \secref{alterbbn}.

\section{Background} \label{sec:background}
In order to form bound nuclei it is necessary to have free neutrons and protons with temperatures below their binding energy. The decoupling of the weak interaction and consequently the appearance of free neutrons and protons happens already around one second after Big Bang, but the binding energy of deuterium is as low as 2.2\,MeV. Since deuterium is the simplest element you can form, the~consequence of the low binding energy is a delay of the nucleosynthesis until 2--3 min after Big~Bang. Meanwhile the free neutrons decay, altering the final abundances of deuterium and helium.

During Big Bang Nucleosynthesis only light elements are formed because helium is very strongly bound (binding energy of $28.3$\,MeV) and there exist no stable nucleus with five nucleons. This means you cannot fuse helium and neutrons/protons into heavier elements, providing a severe bottleneck for the formation of heavier elements. Due to the decay of free neutrons, only the ones that become bound in nuclei survive. Most of the produced deuterium is converted to helium, leaving only of the order of $10^{-5}$ deuterium nuclei per hydrogen nuclei.

The formation of the light elements is sensitive to the expansion rate (temperature) and timing of the events e.g., decoupling, and consequently the abundances of the light elements can probe the conditions during the nucleosynthesis.

\section{Observational Deuterium Measurements} \label{sec:deuteriumobs}
Neutral deuterium has absorption features similar to the Lyman series of hydrogen, but with a $82 \kms$ off-set because of the heavier nucleus. Consequently we can measure the ``fingerprint'' of deuterium caused by absorption in gas clouds along the line of sight to distant bright sources such as quasars. However, due to the fairly small off-set and low abundance of deuterium relative to hydrogen (D/H = $10^{-5}$), high resolution spectra (R $\approx$ 45000) with very good signal to noise are required. Given that there are no significant astrophysical sources of deuterium production \cite{Epstein:1976,Prodanovic:2003} and the destruction rate in stars is low at the relevant redshifts and metallicities \cite{Romano:2006,Dvorkin:2016}, medium-to-high redshift measurements will be very close to the primordial value. 

\begin{figure}[H]
\centering
\includegraphics[width=\columnwidth]{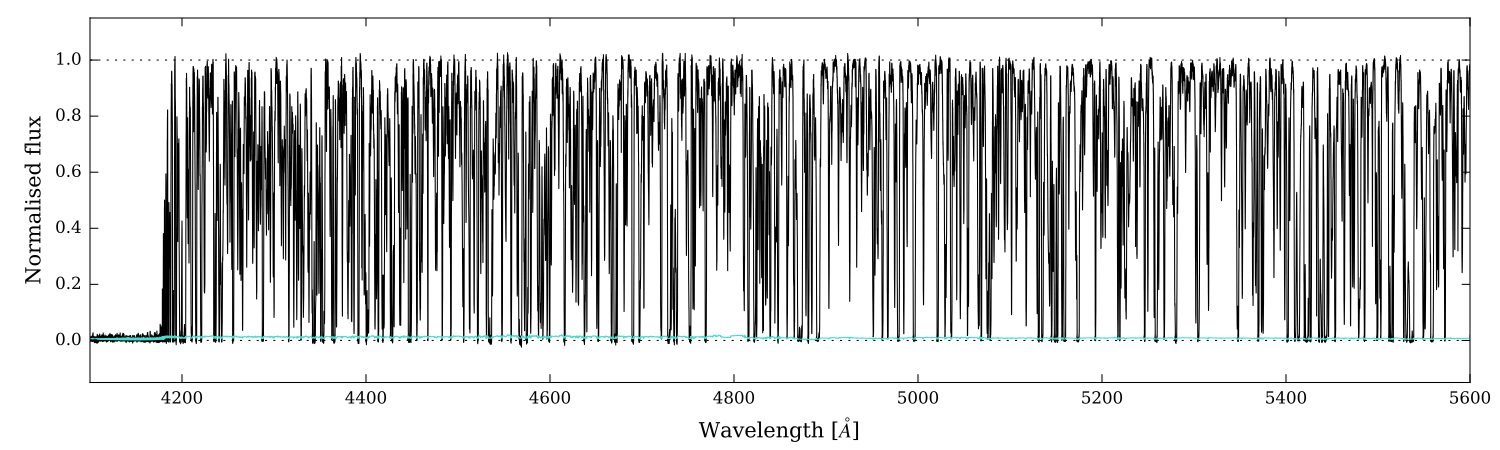}
\caption{The normalised flux spectrum of PKS1937--101 (weighted stacking of all exposures for illustration only). The \Lya{} of the $z_\mathrm{abs} = 3.256$ absorber lies at $5171.7$\,\AA{} and the Lyman Limit at $3877.2$\,\AA{} (outside the spectrum). For the $z_\mathrm{abs} = 3.572$ the \Lya{} is found at $5555.7$\,\AA{} and the Lyman Limit at $4165.1$\,\AA{}.}
\label{fig:spectrum}
\end{figure}

Here we present two new high precision measurements along the line of sight to the quasar PKS1937--101. The absorber at $z_\mathrm{abs} = 3.572$ was previously analysed by \citet{Tytler:1996, Burles:1998a}, and the $z_\mathrm{abs} = 3.256$ absorber was analysed by \citet{Crighton:2004}. New spectra of the quasar have become available in the form of $63.4$\, ks
exposures from the High Resolution Echelle Spectrometer (HIRES) at the Keck Telescope and $32.4$\,ks from the Ultraviolet and Visual Echelle Spectrograph (UVES) at the Very Large Telescope (VLT). A root-mean-square weighted stacking of the new spectra is shown in \figref{spectrum}. The spectra are reduced and normalised following standard procedures and the relevant atomic transitions are fitted with a model consisting of several Voigt-profiles that represent the individual sub-components of the absorbers. Each sub-component is characterised by a redshift, a~column density, a temperature and a turbulent broadening. In addition, we allow for uncertainties in the continuum normalisation and calibration of the spectra. The details of the spectra and their analyses can be found in {Riemer-S{\o}rensen} et al. \cite{Riemer-Sorensen:2015,Riemer-Sorensen:2016}, and here we only discuss the results.

\subsection{Deuterium in the $z_\mathrm{abs} = 3.256$ Absorber} \label{sec:lowz}
\begin{figure}[H]
\centering
\includegraphics[width=\columnwidth]{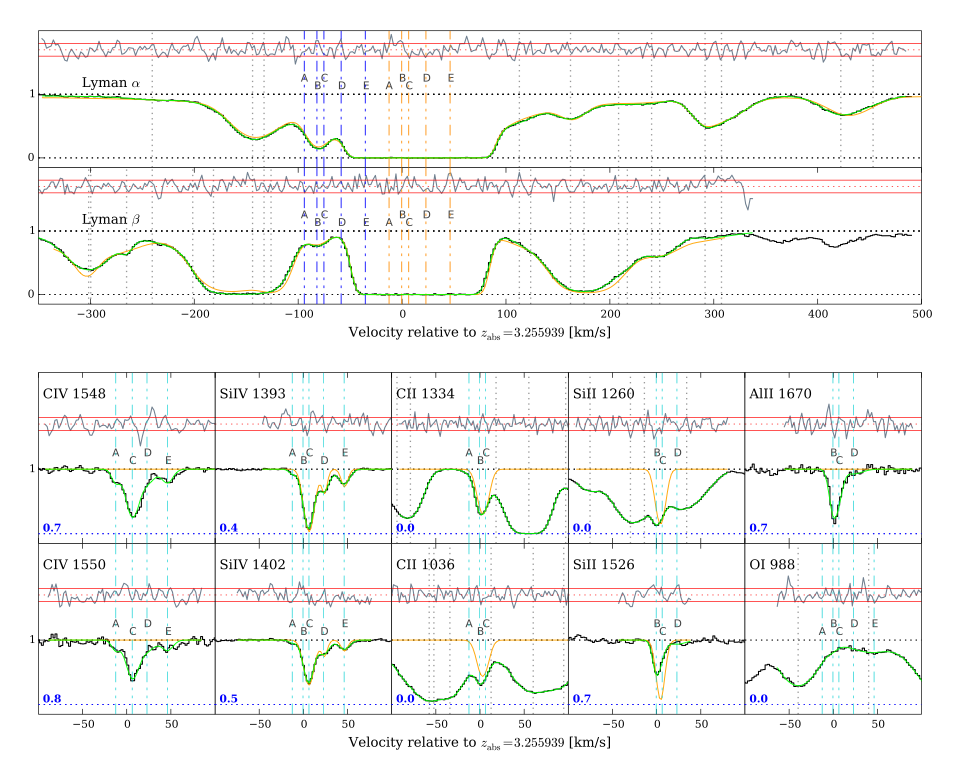}
\caption{The normalised stacked flux spectrum (thick black) of the $z_\mathrm{abs} = 3.256$ absorber, and the best fit model (green), as well as the normalised fit residuals (above each fit). The stacked spectrum is only used for visualisation, as the model fitting is always performed on the individual spectra simultaneously. The fitted velocity components are marked by vertical dot-dashed lines denoted by A~to F (HI in orange, DI in blue, metals in cyan), and interloping HI lines are shown as vertical dotted lines (light grey). It is seen that the structure of the absorber is more complicated than the old model from \citet{Crighton:2004} (orange), which has a significant impact on the final result. Figure from \cite{Riemer-Sorensen:2015}.} 
\label{fig:lowz}
\end{figure}

\textls[-15]{We fit the absorption with Voigt profiles using \vpfit{} \cite{vpfit:2014} to obtain the gas cloud characteristics including the deuterium and hydrogen abundances. We find a deuterium abundance of {\mbox{D/H = \mbox{$2.45\pm0.28\times10^{-5}$}}} to be compared with the previous measurement of D/H = $1.6\pm0.30\times10^{-5}$~\cite{Crighton:2004}.} If naively interpreted as a statistical fluctuation this corresponds to a $3\sigma$ difference. However, it origins in a systematical model difference, which is hard to quantify. The measurement from the new spectra is not more precise despite the increased signal-to-noise ratio, but we believe it to be more robust, as~the better data allow us to resolve the internal structure of the absorber better. This is illustrated in \figref{lowz}. The best fit model is consistent between the UVES and HIRES spectra.

\subsection{Deuterium in the $z_\mathrm{abs} = 3.572$ Absorber} \label{sec:highz}
\begin{figure}[H]
\centering
\includegraphics[width=\columnwidth]{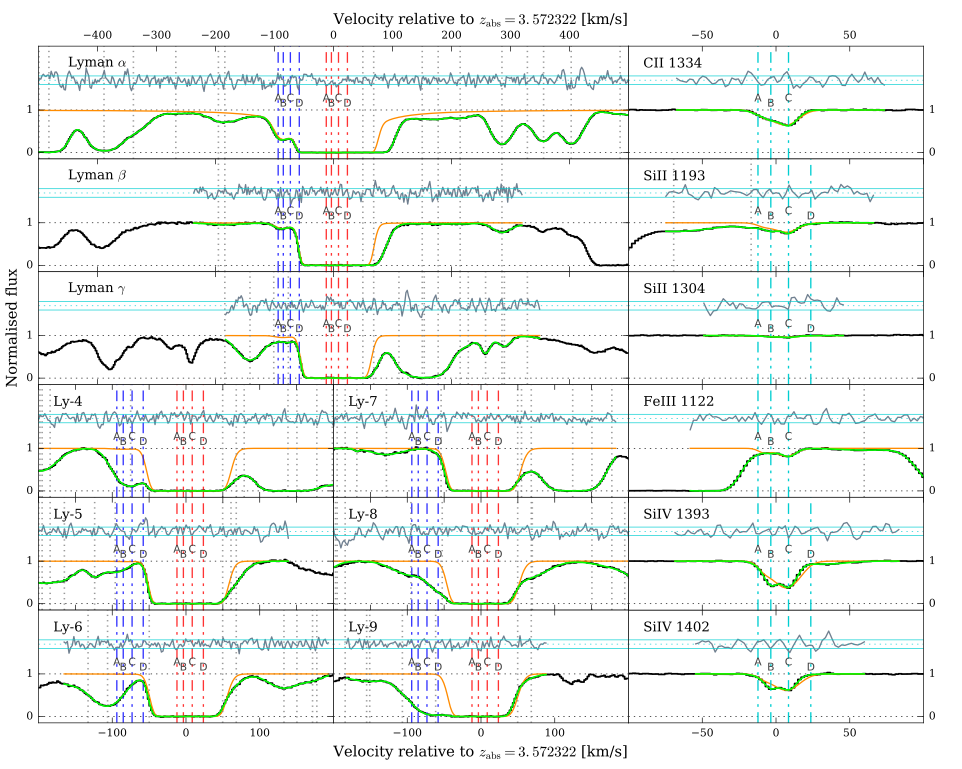}
\caption{The normalised stacked flux spectrum (thick black) of the $z_\mathrm{abs} = 3.572$ absorber, and the best fit model (green), as well as the normalised fit residuals (above each fit). The stacked spectrum is only used for visualisation, as the model fitting is always performed on the individual spectra simultaneously. The fitted velocity components are marked by vertical dot-dashed lines denoted by A to D (HI in red, DI i in blue, metals in cyan), and interloping HI lines are shown as vertical dotted lines (light grey). The model from \citet{Tytler:1996} (without blends) is over-plotted in orange, illustrating the difference between the derived model structures, particularly for the metals. Figure from \cite{Riemer-Sorensen:2016}.}
\label{fig:highz} 
\end{figure}

For the high redshift absorber we find a ratio of D/H = $2.62\pm0.03\times10^{-5}$ to be compared with the previous measurement of D/H = $3.3\pm0.3\times10^{-5}$ \cite{Burles:1998a}. In contrast to \citet{Burles:1998a} we include the heavy element transitions in the model fitting, resulting in a more complicated substructure for the absorber as shown in \figref{highz}. The interesting point here is the very high precision we obtain, even though the hydrogen column density of $\log(N($\Hi$)) = 17.9$ is lower than for previous similar high-precision measurements \cite{Cooke:2014}. This is important for future measurements because the neutral hydrogen column density distribution in quasar absorption systems is a steep power law, with lower column density systems being more common. A statistically large sample of measurements is therefore feasible and at the same time strictly necessary in order to reveal a plateau of primordial values as a~function of, e.g., metallicity. Despite the complexity of the absorber, the very high precision on the D/H ratio is obtained due to a combination of (1) very high signal to noise ratio in the spectra; (2)~fitting 9 Lyman transitions for which the oscillator strengths span more than two orders of magnitude; and~(3)~the presence of several heavy element species that allow us to constrain the velocity structure of the~system.

\subsection{Deuterium Measurement Sample}
For standard Big Bang Nucleosynthesis the inferred value of the primordial deuterium abundance can be used to infer the baryon density of the universe, $\Omega_\mathrm{b}$. \tabref{measurements} provide a list of current measurements, for which the weighted average is $100\Omega_\mathrm{b}h^2 = 2.17\pm0.024$ which deviates by $1.9\sigma$ from the value of $2.225\pm0.016$ inferred from the measurements of the cosmic microwave background \cite{PlanckXIII:2015}. We consider the two results to be consistent.

We find no correlations between deuterium abundance and redshift, metallicity or hydrogen column density (as tentatively claimed by \citet{Cooke:2016}). Neither do we find any evidence for a~dipole in the deuterium measurements \cite{Riemer-Sorensen:2016}. Such a dipole would be expected if there were variations in fundamental constants like the fine structure constant, the hadronic masses or binding energies~\cite{Berengut:2011}. As an example we consider a dipole with the same direction as the potential dipole in the fine structure constant \cite{Webb:2011,King:2012}, for which the preferred slopes are close to zero with uncertainties larger than the preferred value and consequently consistent both with a small dipole and with no dipole.

\begin{table}[H]
\centering
\caption{The sample of \DtH{} measurements from \citet{Riemer-Sorensen:2016}}\label{tab:measurements} 
\scalebox{0.83}[0.83]{
\begin{tabular}{lccccc}
\toprule

\textbf{Reference}				& \textbf{Absorption} \textbf{Redshift}		&\boldmath{$\log (N($}\bf{\Hi{}))}            	& \bf{[X/H]}            	& \bf{\Di/\Hi}    \boldmath{$[\times 10^{-5}]$               } 	& \textbf{100} \boldmath{$\Omega_bh^2$} \\ 
\midrule
\citet{Burles:1998a}			& 2.504		& $17.4\pm0.07$            	& $-$2.55 Si       	& $4.00\pm0.70$       	& $1.66\pm0.18$    \\
\citet{Pettini:2001}			& 2.076		& $20.4\pm0.15$            	& $-$2.23 Si       	& $1.65\pm0.35$       	& $2.82\pm0.36$    \\
\citet{Kirkman:2003}			& 2.426		& $19.7\pm0.04$            	& $-$2.79 O       	& $2.43\pm0.35$       	& $2.24\pm0.20$    \\
\citet{Fumagalli:2011}			& 3.411		& $18.0\pm0.05$            	& $-$4.20 Si       	& $2.04\pm0.61$       	& $2.49\pm0.05$    \\
\citet{Noterdaeme:2012} $^1$	& 2.621		& $20.5\pm0.10$	& $-$1.99 O       	& $2.80\pm0.80$       	& $2.05\pm0.35$    \\	
\citet{Cooke:2014}, \citet{Pettini:2012}		& 3.050		& $20.392\pm0.003$	& $-$1.92 O       	& $2.51\pm0.05$       	& $2.19\pm0.02$    \\
\citet{Cooke:2014}, \citet{OMeara:2001} 	& 2.537		& $19.4\pm0.01$	& $-$1.77 O       	& $2.58\pm0.15$       	& $2.16\pm0.04$    \\
\citet{Cooke:2014}, \citet{Pettini:2008}		& 2.618		& $20.3\pm0.01$	& $-$2.40 O       	& $2.53\pm0.10$       	& $2.18\pm0.03$    \\
\citet{Cooke:2014}			& 3.067		& $20.5\pm0.01$	& $-$2.33 O       	& $2.58\pm0.07$       	& $2.16\pm0.03$    \\
\citet{Cooke:2014}, \citet{OMeara:2006} 	& 2.702		& $20.7\pm0.05$	& $-$1.55 O       	& $2.40\pm0.14$       	& $2.25\pm0.03$    \\
\citet{Riemer-Sorensen:2015}		& 3.255		& $18.1\pm0.03$	& $-$1.87 O		& $2.45\pm0.28$       	& $2.23\pm0.16$      \\	
\citet{Balashev:2016} $^1$		& 2.437		& $19.98\pm0.01$	& $-$2.04 O		& $1.97\pm0.33$	& $2.54\pm0.26$ \\
\citet{Cooke:2016}			& 2.853		& $20.34\pm0.04$	& $-$2.08 O       	& $2.55\pm0.03$       	& $2.17\pm0.03$    \\
\citet{Riemer-Sorensen:2016}		& 3.572		& $17.925\pm0.006$ 	& $-$2.26 O		& $2.62\pm0.05$	& $2.14\pm0.03$    \\	
\midrule

Weighted average $^1$		& ---		& ---		& ---		& $2.55\pm0.02$	& $2.17\pm0.02$  \\
Unweighted average $^1$		& ---		& ---		& ---		& $2.53\pm0.16$	& $2.18\pm0.08$ \\
\midrule

\citet{PlanckXIII:2015}\			& ---		& ---		& --- 		& $2.45\pm0.05$	& $2.225 \pm 0.016$\\
\bottomrule
\end{tabular}}\\

\begin{tabular}{@{}c@{}} 
\multicolumn{1}{p{\textwidth -.88in}}{\footnotesize  In order to convert between \Di/\Hi{} and $\Omega_bh^2$ we assume standard nucleosynthesis and use the nuclear rates from \citet{Coc:2015}. $^1$ The \citet{Balashev:2016} and \citet{Noterdaeme:2012} measurements are excluded from the average because they are derived under the assumption of constant \Oi/\Hi{} across all components, which may not be an appropriate assumption for a high precision measurement.}
\end{tabular}
\end{table}

\section{Predictions from Nucleosynthesis} \label{sec:alterbbn}
The improved observational precision calls for nucleosynthesis predictions with similar accuracy. We have updated and expanded the existing AlterBBN code with nuclear rates and non-standard physics options. AlterBBN is a {publicly available} C code for evaluating abundances \cite{Arbey:2012}, which is based on the Wagoner code \cite{Wagoner:1969} and similar to NUC123 (\cite{Kawano:1992}, also known as the Kawano code)  and PArthENoPE \cite{Pisanti:2008} \footnote{PArthENoPE is also publicly available, but it requires expensive fortran libraries to compile}. {AlterBBN can be run as an independent code or included in multi-parameter analyses using e.g., CosmoMC \footnote{\url{http://cosmologist.info/cosmomc}} or Montepython \footnote{\url{http://baudren.github.io/montepython.html}}}. Until {the improvements described here} are implemented in the official version of AlterBBN, the code and further details can be obtained from \url{github.com/espensem/AlterBBN}.

AlterBBN evaluates the {light element} abundances as a function of time during the Big Bang Nucleosynthesis (BBN). It starts from a thermodynamic equilibrium, whereafter the weak interaction and the neutrinos decouple, leading to the decay of free neutrons. Eventually the electrons and positrons will decouple and annihilate, which will reheat the photons (and any other particles with an electromagnetic coupling). Until the photon temperature has dropped below the binding energy of deuterium, the~protons and neutrons cannot form any deuterium.

The resulting abundances depend on the nuclear reaction rates and the timing/expansion in the early universe. Without any non-standard physics, the baryon to photon ratio is the only free parameter in BBN. Since it can be determined very precisely from the cosmic microwave background~\cite{PlanckXIII:2015}, the~main uncertainty in the predictions comes from the nuclear reaction rates.

\subsection{Updating Nuclear Reaction Rates}
We have extended the nuclear network in AlterBBN from 88 to 100 reactions, and updated six important rates with new theoretical/experimental determinations \cite{Coc:2015}. The resulting helium and hydrogen-3 abundances are relatively unaffected by these changes, but D/H change by 4.5\%, and the uncertainty double. Lithium-7 and beryllium-7 increase by 18\% thereby worsening the lithium~problem.

An example of the updated rates is the formation of helium-3 via d(p,$\gamma$)$^3$He. For this rate there is only very scarce experimental data at the relevant energy range \citep{Ma:1997}. Instead the rates have been determined theoretically or from fitting various polynomials or theoretical models to an~extended energy range and extrapolating to the range of interest. The newest theoretical calculation \mbox{by \citet{Marcucci:2016}} is inconsistent with the experimental rate measured around 0.1~MeV. To~reconcile the discrepancy \citet{Coc:2015} rescaled the theoretical model from \mbox{\citet{Viviani:2000,Marcucci:2005}} to the experimental rates for the full range of 0.002--2 Mev. Using AlterBBN with the theoretically determined rate from \cite{Marcucci:2016} leads to D/H = $2.49\pm0.03\pm0.03\times10^{-5}$ (the two errors are due to nuclear rate uncertainties and uncertainties in the baryon density, respectively) while the experimental rescaling from \cite{Coc:2015} leads to D/H = $2.45\pm0.057\times10^{-5}$.
{The two results deviate by approximately $1\sigma$. 

Another option is to use the experimental data directly (as in e.g., \cite{Iocco:2009,Cyburt:2016}) \footnote{To be implemented in the next version of AlterBBN}, which will further increase the deviation between the various predictions from AlterBBN. At this point we do not want to advocate one choice over another, but simply want to emphasise that the choice of rates matters, and unless the specific choice is discussed in the context of each result (which is usually not the case (e.g.,  \cite{Cooke:2014,Riemer-Sorensen:2015,Balashev:2016})), this deviation between rates should be included as a systematic uncertainty. Better~experimental determination of the rates at BBN energies will solve the issue.}

\subsection{Implementing New Physics}
Non-standard physics such as extra relativistic species, modified gravity or lepton asymmetry may change the expansion rate in the early universe or the order of the events governing the nucleosynthesis~\cite[]{Steigman:2012,Nollett:2014,Nollett:2015}. These effects could potentially origin in production/destruction of dark matter or neutrino-like species. At present and, likely for the foreseeable future, BBN provides the only window to a universal lepton asymmetry. The most obvious way to generate lepton asymmetry is via additional neutrino species

We have updated AlterBBN with the possibility of adding the presence of equivalent neutrinos (non-interacting relativistic species) and Weakly Interacting Massive Particles (WIMPs) during the nucleosynthesis.  The WIMPs are generic and their effect is purely determined by their mass, the~number of internal degrees of freedom, their nature (fermion or boson) as well as any coupling to the standard model particles. The WIMPs can be either electromagnetically coupled or coupled to neutrinos (standard model neutrinos and any equivalent species) (following \cite{Nollett:2014,Nollett:2015}). They are not required to provide 100\% of the dark matter but could be a subcomponent. We consider four types of WIMPs:
(1) Real scalars that are self-conjugate and have one degree of freedom;
(2) Complex scalars that have two degrees of freedom;
(3) Majorana fermions which are self-conjugate and have two degrees of freedom; and
(4) Dirac fermions with four degrees of freedom.

\begin{figure}[H]
\centering
\includegraphics[width=0.35\columnwidth]{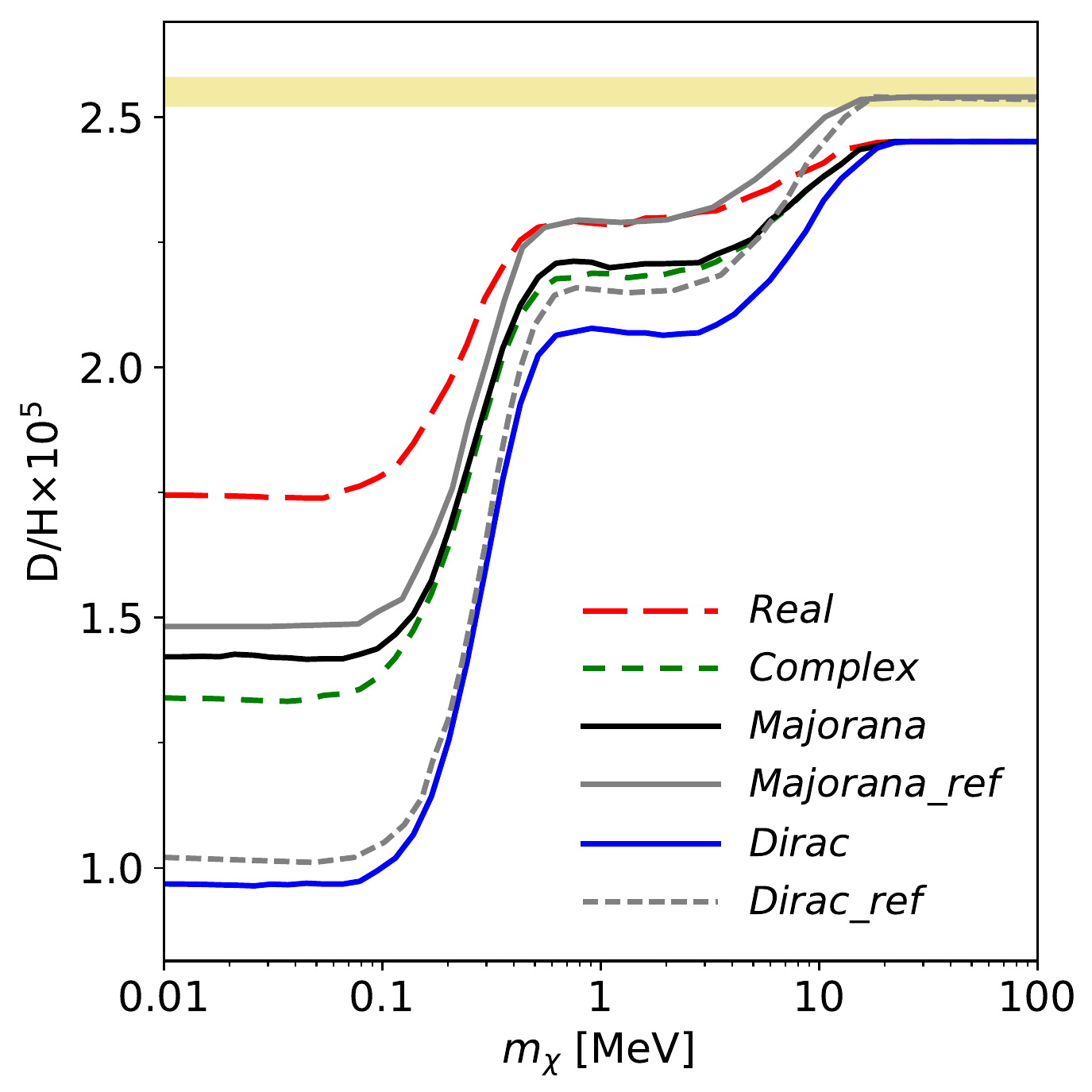}
\includegraphics[width=0.35\columnwidth]{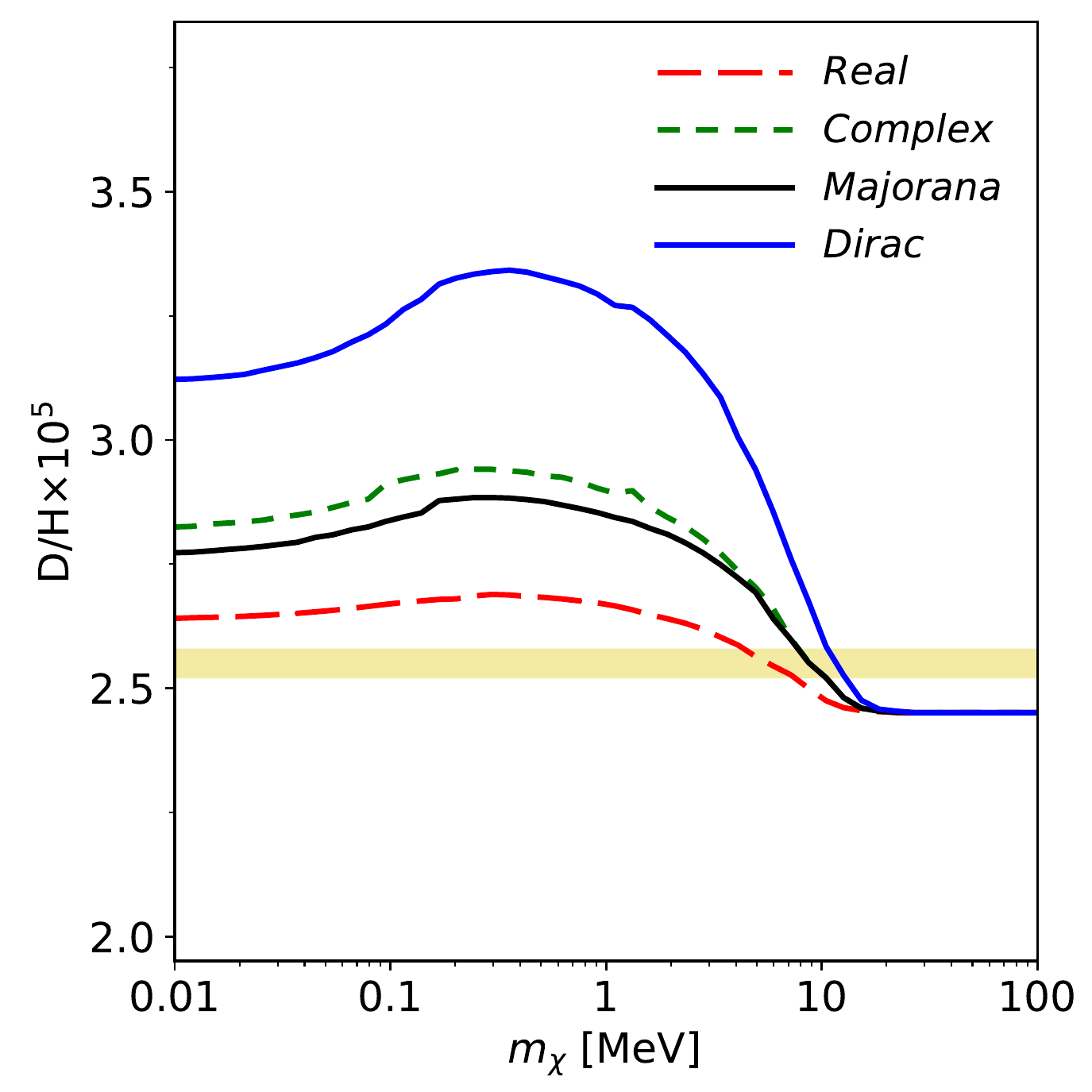}
\caption{\textls[-5]{The deuterium abundance as a function of Weakly Interacting Massive Particles (WIMP) mass if the WIMP is a real scalar (dashed~red), a complex scalars (dotted green), Majorana fermion (solid black), Dirac fermion (solid~blue). The left plot is for electromagnetically coupled WIMPs (coupling to electrons or photons) and the right plot is for neutrino coupling. The grey lines are for comparison to \cite{Nollett:2014,Nollett:2015}. The deviations are due to different choices of nuclear rates. The yellow horisontal bands are the observed ratio of D/H \cite{Riemer-Sorensen:2015}. \mbox{Figure from \cite{Jenssen:2016}}.}} 
\label{fig:mX_DH}
\end{figure}

The presence of such non-standard model particles will alter the expansion rate because they contribute to the total energy density. They might also increase the temperature of photons or neutrinos due to annihilation, and by mimicking neutrinos, they might change the inferred number of relativistic species {often parametrised as a change in the number of neutrino species $N_\mathrm{eff} = 3.046 + \Delta N_\nu$ where $\Delta N_\nu$ is the equivalent number of neutrinos that would provide the same effect (e.g., \cite{Lesgourgues:2006,Riemer-Sorensen:2013}). The~relativistic particles contribute to the radiation energy density so that $\rho_\mathrm{rad} \rightarrow \rho_\mathrm{rad}' = \rho_\mathrm{rad}+\Delta N_\nu \rho_\nu$. Combining with the Friedman equation \cite{Friedmann:1924} we can relate the expansion factor directly to $N_\mathrm{eff}$ as }
\begin{equation}
\frac{H'}{H} = \left(\frac{\rho_\mathrm{rad}'}{\rho_\mathrm{rad}} \right)^{1/2} = \left(1+\frac{7 \Delta N_\nu}{43} \right)^{1/2}
\end{equation}
{where $H'/H$ is the relative expansion rate with and without the extra species. Similarly, lepton asymmetry can be parametrised in terms of $N_\mathrm{eff}$ (see e.g., \cite{Steigman:2012}), and an increase in entropy (e.g.,~from decaying dark matter) will affect the number of degrees of freedom, which can also be expressed as a~change in $N_\mathrm{eff}$. 

In \figref{etaVSNeff} we show the probability contours for simultaneously varying the baryon density ($\eta=n_\mathrm{b}/n_\gamma$) and the additional number of neutrinos for otherwise standard BBN and comparing to observational constraints on \DtH{} and $^4$He. As seen the constraints from BBN alone are consistent with those from \Planck{} and of comparable precision. The combined \DtH{} and $^4$He constraint on $\Delta N_\nu$ is $\Delta N_\nu = 0.06^{+0.27}_{-0.36}$.

\begin{figure}[H]
\centering
\includegraphics[width=0.35\columnwidth]{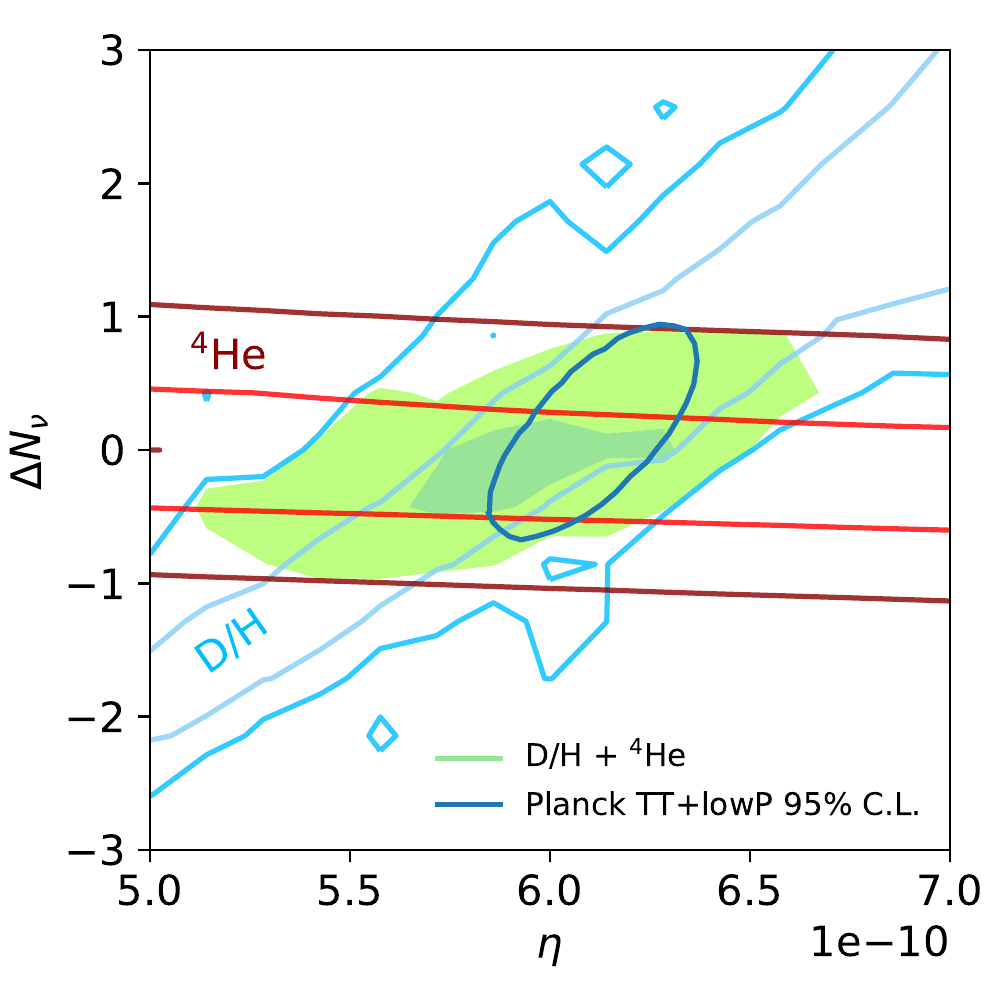}
\caption{Constraints in the $\eta-\Delta N_\nu$ plane (68\% and 98\% C.L.) from observations of $D/H$ (light blue lines, from \citep{Riemer-Sorensen:2016}), $^4$He (red lines, from \citep{Aver:2015}), and combined (green shade). We also show the 95\% contours for \Planck{} for comparison (dark blue, from TT + LowP in \citep{PlanckXIII:2015}).}
\label{fig:etaVSNeff}
\end{figure}   

Including WIMPs will change the radiation and/or matter energy density and pressure directly, as well as possible changes in the entropy and lepton asymmetry depending on the specific properties. The expansion rate governs the particle number densities and consequently the interaction rates for nucleosynthesis.} \figref{mX_DH} illustrate the changes of the D/H ratio when including WIMPs in the mass range from 0.01 to 100 MeV (heavier or lighter WIMPs will provide linear extensions of the abundance in \figref{mX_DH} as they decouple completely before/after the nucleosynthesis). The results are compared to those of \cite{Nollett:2014,Nollett:2015}. The deviations are due to different choices of nuclear rates (e.g., the \cite{Nollett:2014,Nollett:2015} results are based on the theoretical calculation of d(p,$\gamma$)$^3$He from \cite{Viviani:2000} while we have used the experimentally rescaled rate from \cite{Coc:2015}). Comparing to the observational constraints, we see that for electromagnetically coupled WIMPs we can exclude masses below 10 MeV and for neutrino coupled WIMPs only masses around 10 MeV are allowed.

Neither non-standard physics nor the new rates alleviate the Lithium problem.

\section{Summary}
We have presented new high precision measurement of the deuterium to hydrogen ratio from low column density absorbers ($N($\Hi$)= 17.9$ and $18.1\cm^{-2}$) along the line of sight to the quasar PKS1937--101.  This is important for future measurements because the neutral hydrogen column density distribution in quasar absorption systems is a steep power law, with lower column density systems being more common. A statistically large sample of measurements is therefore feasible. 

With the acquired precision on the observational measurements, it is timely to similarly improve the theoretical predictions from Big Bang nucleosynthesis. We have updated the publicly available code, AlterBBN, with new reaction rates and non-standard physics. It is clear that the choice  between theoretical and measured rates at relevant energies leads to different results, providing a source of systematic uncertainty which is often neglected.

\vspace{6pt} 

\supplementary{{The updates of AlterBBN is available here \url{github.com/espensem/AlterBBN} and will be included in the next official and distributed release of AlterBBN.}}

\acknowledgments{The authors would like to thank the anonymous referees for constructive comments and~suggestions.}

\authorcontributions{All the authors equally contributed to this work.}
\conflictsofinterest{The authors declare no conflict of interest.}

\bibliographystyle{mdpi}
\renewcommand\bibname{References}

\end{document}